# Cosmic Rays from Supernova Remnants and Superbubbles

Richard E. Lingenfelter[*]
Center for Astrophysics and Space Sciences, University of California San Diego,
La Jolla, CA, USA.[*] rlingenfelter@ucsd.edu; (858)-534-2464

**Abstract.** Recent high energy gamma-ray observations of both single supernova remnants and superbubbles, together with observations of supernovae, star formation regions, and local cosmic ray composition, now provide an integrated framework tying together the sources, injection, acceleration, and propagation of the cosmic rays, so that it is possible to determine their relative contributions to cosmic ray acceleration for all but the very highest energies.

1. Introduction

   Recent *Fermi* and other observations (Ackermann et al., 2011, 2013, 2016; Acero et al., 2016) of high energy gamma rays from the decay of neutral pions produced by cosmic rays have at long last directly confirmed their shock acceleration both in single supernova remnants and in superbubbles. These measurements are also broadly consistent with the simple power-law energy spectra predicted in test particle models of diffusive shock acceleration (e.g. Blasi, 2013). But new measurements (Ahn et al., 2010; Aguilar et al., 2015, 2016) of the local cosmic ray nuclear spectra reveal small, yet very significant deviations from such a simple spectra in the form of a hardening of the He and C spectra with respect to protons at high energies not expected from diffusive shock acceleration (e.g. Ohira et al., 2016).
   Further measurements (Rauch et al., 2009; Binns et al., 2016; Murphy et al., 2016) of the spectrum and nuclear composition of the local cosmic rays, also reveals their source composition together with the combined effects of their energy dependent escape and propagation through the interstellar medium and better define their injection processes and acceleration times. In addition, new studies of the nature of Galactic supernovae now allow us to determine the relative contributions of their two major sites of supernova shock acceleration: isolated supernova explosions in the broad interstellar medium and collective explosions in the superbubbles that they create around OB associations.

   Supernovae have long been recognized (e.g. Baade & Zwicky, 1934) as the most powerful known sources of explosive energy in our Galaxy. Estimated (Li et al., 2011) to occur at an average rate of one supernova every ~35 years, releasing about $10^{51}$ ergs in their ejecta and shocks (e.g. Nomoto et al., 1984; Woosley & Weaver, 1995), they generate an average supernova explosive power of $\sim 10^{42}$ ergs/s. This is roughly an order of magnitude greater than the $\sim 10^{50}$ ergs/SN that observations have shown (e.g. Lingenfelter, 1992, 2013) are needed to maintain the present Galactic cosmic ray luminosity.

Extensive studies (e.g. Axford, 1981; Bykov & Fleishman, 1992; Ellison et al., 1997; Blasi, 2013) all suggest that strong diffusive shock acceleration in the expanding supernova remnants is the most effective process, generating the cosmic rays with ~10% efficiency. Such strong shocks with compression ratios of 3 to 4 are expected to produce power-law spectra at relativistic energies with an index of -2 to -2.5. This range is quite consistent with the cosmic ray proton and electron spectral indexes of around -2.2, implied by supernova synchrotron radio remnants and the recent high energy gamma rays produced by cosmic ray proton interactions in both isolated supernova remnants and collectively in superbubbles.

Moreover, that source spectral index is in excellent agreement with the values of -2.2 to -2.4, determined from the local cosmic ray proton and nuclear index of -2.7 (e.g. Engelmann et al., 1990; Obermeier, et al., 2012) after correcting it by -0.33 to -0.5 for the energy dependent cosmic ray escape lifetime in the Galaxy. That spectral index correction, reflecting a steepening of the source spectrum during propagation was determined from measurements (Obermeier et al., 2012; Adriani et al., 2014; Aguilar et al., 2016) of the spallation production ratio of secondary B to parent CNO nuclei as a function of energy.

## 2. Single Supernova Remnants in the Interstellar Medium

Protons make up ~92% of the cosmic rays and carry ~73% of their total mass and energy, while helium and all of the other heavier nuclei, respectively, make up just ~8% and 0.4% of the particles with ~17% and ~7% of the energy (Cummings et al. 2016). But ironically it was the cosmic ray electrons and positrons, which amount to only ~1% of the particles and only carry ~4% of the energy, that were first shown (e.g. Ginzburg & Syrovatskii, 1964) to be accelerated by supernova shocks. Single supernovae, exploding randomly throughout the Galaxy in the predominantly warm ~$10^4$ K, dense ~1 H/cm$^3$ phases of the interstellar medium, turn on as radio synchrotron emitting remnants after just a few hundred years, as relativistic electrons are accelerated in their shocks and radiate by synchrotron emission at radio frequencies in the surrounding shell of swept up, compressed and amplified magnetic field (e.g. Gull, 1973; Ohira et al., 2012). This synchrotron emission, clearly identifying the acceleration of relativistic cosmic ray electrons, lights up such supernova remnants with a surface brightness that allows them to be seen by radio telescopes throughout much of the Galaxy, and roughly 300 have been cataloged since the 1960s (Green, 2014).

Now, the long awaited proof of cosmic ray proton acceleration in single supernova remnants has at last been found (Giordano et al., 2012; Ackermann et al., 2013) in the *Fermi, VERITAS* and other high energy gamma ray spectra of both a small number of very young thermonuclear SN Ia in the general interstellar medium and the more numerous slightly older core collapse supernovae interacting with parts of their parent molecular clouds. The combined spectral range of these measurements from a GeV to ~10 TeV was wide enough to show the distinctive signature of the two-gamma decay of neutral pions in their highly relativistic rest frame produced by proton-proton interactions, as well as the roughly -2 to -2.5 power law spectrum expected from strong shocks in the test particle models of diffusive shock acceleration.

The first measured of these remnants was fittingly that of Tycho's first telescopic supernova of 1572, a roughly spherical shell in radio at a distance of 2.8 to 3.5 kpc, and close to the detection threshold of *VERITAS* at a TeV, which limited the spectral sensitivity. The first analysis (Morlino & Caprioli, 2012) of the gamma ray spectrum (Acciari et al., 2011; Giordano et al., 2012), Figure 1, shows that neutral pion decay gamma rays from cosmic ray proton interactions provide a good fit, as do later analyses (e.g. Berezhko et al., 2013; Slane et al., 2014), while electron bremmstrahlung and Inverse Compton models do not. Updated data (Park et al., 2015) from *Fermi* and *VERITAS* with twice the observing time extend both to slightly lower energies, adds an upper limit of 0.5 from 100-300 MeV, clearly showing the pion decay signature below 200 MeV, and reveals a steepening of the spectrum to a -2.9±0.4 power law above ~0.5 TeV, but leaves the spectrum in between unchanged.

Over that long intermediate span from ~500 MeV to ~500 GeV the gamma ray energy spectrum is well fitted (Morlino & Caprioli, 2012) by a single power of -2.1 to -2..2, reflecting the cosmic ray acceleration spectrum up to proton energies of ~50 TeV. This power law is only slightly steeper than the -2.0 expected for the strongest shocks in simple diffusive shock acceleration. But it is significantly different from the concave spectra expected from standard nonlinear diffusion shock acceleration, which is even flatter than -2.0 (e.g. Blasi, 2013 for reviews). Later models have appealed to environment effects to save the -0.2 spectrum by adding a second low energy component from either denser nearby clumps of gas (Berezhko et al., 2013) or Inverse Compton (Slane et al. 2014) from a more intense radiation field, including infrared from the Tycho remnant itself. Over all, these observations also imply that fully 12 to 16% of the supernova ejecta energy goes into cosmic rays.

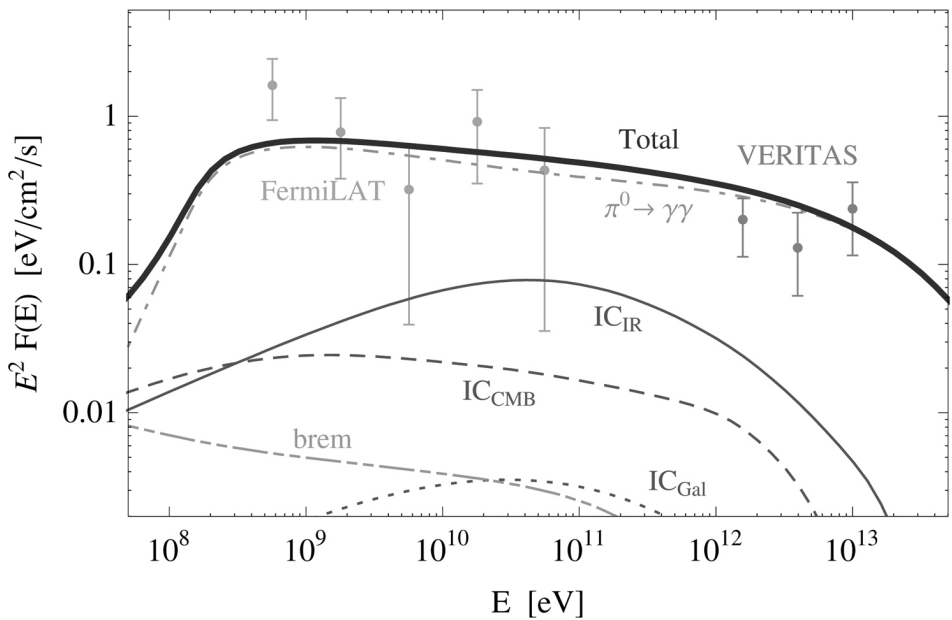

Figure 1. The gamma ray spectrum times $E^2$ from the young thermonuclear SN Ia remnant discovered Tycho Brahe in 1572, measured by *Fermi* (Giordano et al., 2012) and *VERITAS* (Acciari et al., 2011), and compared to both lepton and hadron emission models (Morlino & Caprioli, 2012), clearly fits the pion-decay gamma ray signature from cosmic ray proton interactions rising up to ~200 MeV, then "flattening" to ~$E^{-2}$ as expected from diffusive shock acceleration all the way up to about a TeV before starting to break to a steeper escape spectrum, while electron bremmstrahlung and Inverse Compton models don't come close.

So far gamma ray spectra have been measured from only two other SN Ia remnants, SN 1006 and SN 185 (RCW 86), which are about 2 and 4 times as old as Tycho, but still among the youngest remnants. Their numbers are consistent with the estimated occurrence rate of ~1 SN Ia/180 yr in our Galaxy (Li et al., 2011). These also have slightly different spectra and have been fitted by both pion decay and Inverse Compton gamma rays, or likely a combination there of reflecting further environmental variations. So even though SN Ia explosions are "standard candles," they are randomly born in diverse environments and the relative mixes and efficiencies of their cosmic ray lepton and hadron acceleration and interactions are quite variable (e.g. Ohira et al., 2012).

Far more radical environment conditions have also produced a few dozen single supernova remnants from the much more frequent core collapse supernovae that also show the characteristic gamma ray spectrum of cosmic ray proton production. The first of these were two well-studied "middle-aged" ~10,000 year-old, ~20 pc diameter, radio remnants IC443 and W44 at distances of 1.5 and 2.9 kpc (Figure 2). Unlike the common radio shell remnants, these two belong to a much more complex class of "mixed-morphology" remnants (Rho & Petre, 1998), whose markedly different radio and X-ray images together with other characteristics suggest that they are interacting with nearby molecular clouds, observed at other frequencies.

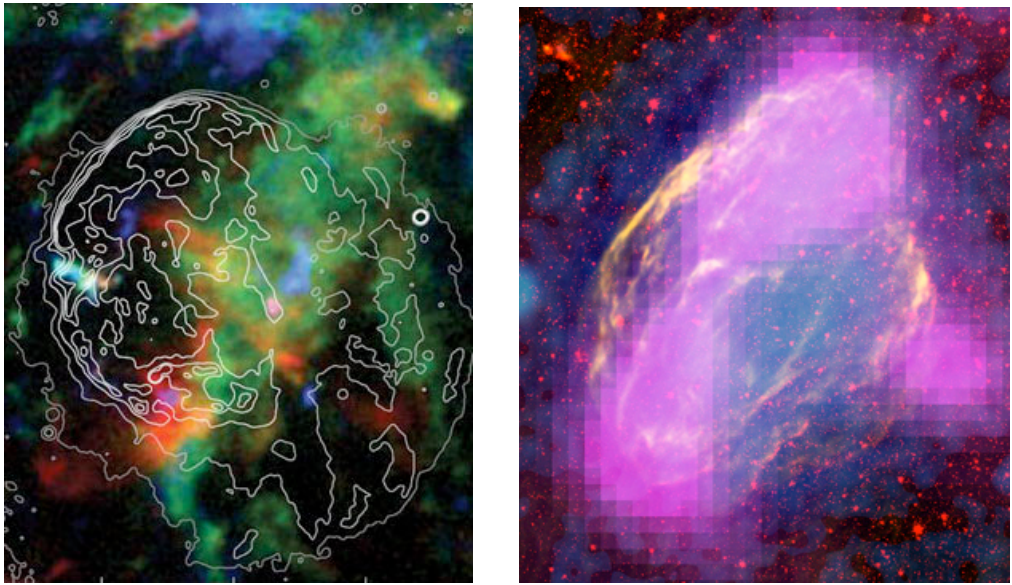

Figure 2. Multi-wavelength images of supernova remnants IC443 (left) and W44 (right), whose pion-decay gamma rays have been detected by *Fermi* (Ackermann et al., 2013) from freshly accelerated cosmic ray protons interacting in the dense, shock "crushed" gas of nearby filamentary molecular clouds and clumps, which are shown here in $^{12}CO$ (color) with the 21-cm radio contours of IC443 remnant superimposed, and are shown directly in the high energy gamma ray emission (pixelated) superimposed over a composite image of radio, infrared and X-ray remnant W44. (Lee et al., 2015; NASA Fermi LAT Collaboration)

Both of their gamma-ray flux spectra clearly show (Ackermann et al., 2013) the proton produced pion decay feature as a rising segment below ~200 MeV in Figure 3, again multiplied by $E^2$. Above that energy the spectrum breaks to a broad maximum, corresponding to the proton acceleration spectrum before steepening to the escape spectrum. The modeled proton spectra responsible for these gamma rays are also shown with a best-fit power-law "acceleration" index of -2.4 breaking to an "escape" index of -3.1 above ~240 GeV in IC443, assuming cosmic ray interaction in dense 20 H/cm$^3$ gas in part of a molecular cloud, and a similar best-fit index of -2.4 breaking to -3.5 above ~22 GeV in W44 in 100 H/cm$^3$ gas. These and subsequent spectral measurements of W51C and roughly a dozen other young, interacting remnants, mostly <1,000 years old, also show (Acero et al. 2016: Jogler & Funk, 2016) the pion-decay signature of protons, clearly supporting a coherent model of diffusive shock acceleration of cosmic ray protons in single supernova remnants interacting in molecular gas.

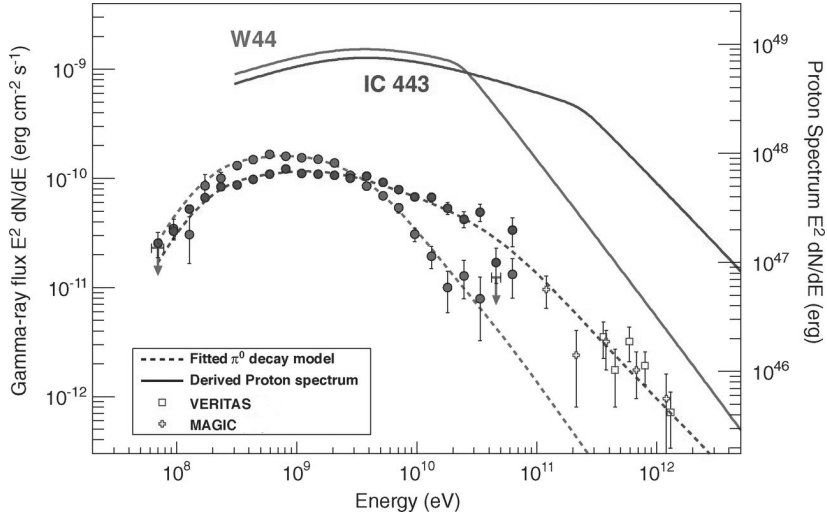

Figure 3. The flux spectra times E$^2$ of the ~100 MeV to ~1 TeV gamma rays from the supernova remnants W44 and IC433, measured by *Fermi* (Ackermann et al., 2013), *VERITAS* (Acciari et al., 2009) and *Magic* (Albert et al., 2007). These spectra are consistent with cosmic ray proton interactions, showing their identifying pion-decay signature below ~200 MeV, then briefly "flattening" to ~E$^{-2}$ as expected from diffusive shock acceleration and breaking to steeper escape spectra above ~2 GeV and ~22 GeV respectively. The infered proton spectra (solid lines) that break at higher energies are also shown (from Ackermann et al., 2013).

The supernova remnants interacting in molecular clouds are, in fact, the dominant class of the observed Galactic high energy gamma ray remnants. For the *Fermi* group (Acero et al., 2016) has now uniformly surveyed all of the known radio supernova remnants (Green, 2014) looking for more such sources in the gamma ray energy range from 1 to 100 GeV. Though they found just 30 gamma-ray supernova remnants that they classified as "likely GeV SNRs," with a possible 14 marginal candidates, or just 10-15% of the 289 radio remnants surveyed, they pointed out that the gamma ray emission from each of their likely supernovae were consistent with their producing the estimated ~10$^{50}$ ergs/SN that would be needed for all Galactic supernovae to produce the present Galactic cosmic rays. They also noted that only these supernova remnants interacting with dense molecular clouds were bright enough to be detected above current thresholds. Some have taken these observations as evidence that solitary supernova remnants in the interstellar medium are the major source of Galactic cosmic rays. But there is a fundamental problem with that. All ~300 of the known supernova radio remnants in the interstellar medium were produced by only a fraction of all the Galactic supernovae (Clark & Caswell, 1976; Higdon & Lingenfelter 1980).

### 3. Clustered Supernovae in Their Superbubbles

Most supernovae do not occur randomly throughout the interstellar medium. Instead, their progenitors like most of the stars, of which they account for just ~0.1%, are born in episodic bursts of many thousands of stars in a Myr in tight clusters of only a few pc in diameter in the densest parts of giant molecular clouds that contain up to ~$3\times10^6$ $M_\odot$ of gas and dust (e.g. Pudritz, 2002). The most massive of these, the 8 to 120 $M_\odot$, most rapidly evolving stars, the O 15-120 $M_\odot$ (Martins et al., 2005) and the B below, end their lives as core collapse supernovae. Moreover, they pass on much of their natal stellar clustering from their compact OB associations. With a mean radial dispersion velocity of only ~2 km/s, or ~2 pc/Myr (de Zeeuw et al., 1999), most of these supernova progenitors disperse <50 pc (Higdon et al., 1998; Higdon & Lingenfelter, 2005) during their relatively short lives of just 3 to 35 Myr before they become supernovae (Schaller et al. 1992; Chieffi & Limongi 2013). Thus, the core-collapse supernovae are also *highly correlated in both space and time*. Such core collapse supernovae account (Li et al., 2011) for ~81% of the Galactic total at a rate of 1 every ~43 yrs. That leaves the thermonuclear SN Ia from far more slowly evolving, ~ Gyr or more (Greggio, 2005), and hence widely scattered, accreting white dwarfs to produce the bulk of single supernova remnants in the interstellar medium at 1 every ~180 yrs.

There are usually under 100 supernova progenitors in a typical OB association, which observations (e.g. Blaauw, 1991) show are generally produced as part of a series of several bursts of star formation separated in space and time by ~50 pc and ~4 Myr, as their parent molecular cloud complexes form multiple OB associations. The bigger bubbles around these close OB associations frequently coalesce into an even larger superbubble. As few as 5 supernovae in a cluster appear to be all that is needed to generate a superbubble (Higdon & Lingenfelter, 2005). So from the size distribution of the young star clusters, ~85% of the core collapse stellar progenitors are expected to be born in superbubble generating clusters. But in that crowded environment, as much as half of those smaller OB clusters may also be engulfed by neighboring superbubbles.

In addition, it appears that essentially all of that fringe of core collapse supernova progenitors that miss being engulfed by superbubbles, still explode very close to some part of the parent molecular cloud complex and can account for essentially all of those "interacting" gamma ray supernova remnants observed by *Fermi* in the interstellar medium. For, assuming a detectable lifetime (Acero et al., 2016) of anywhere from ~10 to 50 kyr, those 30 to 44 detected plus marginal gamma ray remnants interacting with molecular gas, give an occurrence rate of ~1 to 4 SNR/kyr out of ~23 SNR/kyr for all galactic core collapse supernovae. Independent of the *Fermi* sky coverage, this amounts to ~4 to 17% of these core collapse supernovae, and from a quarter to all of the estimated ~15% that were born in some of the smaller clusters that were unable to create superbubbles of their own.

Thus we find (Higdon & Lingenfelter, 2005) that core collapse supernovae in superbubbles likely account for ~75% of all Galactic supernovae, occurring at an overall rate of 1 SN every ~50 years with an average of ~300 supernova occurring in each of some ~3,000 merged superbubbles over their ~50 Myr lifetimes. The supernovae in these superbubbles, do not produce many, small ~20 pc, individual supernova remnants like those in the warm, dense phases of the interstellar medium that radiate away most of their shock energy in <50 kyr. Instead the bursts of supernovae resulting from these highly clustered stars that merge into single, much larger superbubbles, can grow to as much as a few hundred pc as the bubbles of nearby associations merge and they continue accelerating cosmic rays for ~50 Myr or more until their last supernova explodes.

The powerful Wolf Rayet winds of the most massive O stars, M >25 $M_\odot$, and their subsequent supernovae blew cavities in the molecular clouds and swept up the surrounding interstellar gas and magnetic field into warm ~$10^4$ K, dense ~10 H/cm$^3$, HI shells, or cocoons. These enclosed their merged supernova remnants in giant superbubble cavities of >$10^5$ pc$^3$ filled with hot >$10^6$ K, tenuous ~$10^{-2}$ H/cm$^3$ gas (e.g. McKee & Ostriker, 1977; Weaver et al., 1977; Mac Low & McCray, 1988; Tomisaka, 1992; Shull & Saken 1995). Because of the asymmetry of the swept up magnetic field pressure, the superbubbles tend to be roughly tubular in shape with the long axis along the magnetic field.

These roughly 3,000, hot, tenuous, cosmic ray accelerating superbubbles, each filling close to ~$10^6$ pc$^3$, occupy a total of about $3 \times 10^9$ pc$^3$, which gives them a filling factor of just ~1% of the Galactic disk volume of ~$5 \times 10^{11}$ pc$^3$. After their supernovae cease, the hot, supernova-enriched plasma mixes slowly into the previously superbubble-generated hot ionized phase of the interstellar medium of the disk and into the overlying halo, driving the hot phase filling factor up to the recently estimated values of as much as ~50% (e.g. Cox, 2005; Kennicutt & Evans, 2012). That suggests that a significant fraction of the SN Ia which are expected to produce most of the single supernovae in the interstellar medium should occur in the hot, post-superbubble phase, not just in the warm phases as often assumed, making the gamma ray emission from their cosmic ray interaction fainter and even harder to detect.

The thermal X-ray emission from the hot tenuous cores of these superbubbles is generally too faint and diffuse to be resolved in all but some of the closest and largest bubbles (Cash et al., 1980). But the surrounding, much denser HI supershells are readily seen in 21-cm emission, and these and their enclosed voids led to their discovery (Heiles, 1979). The youngest superbubbles are also revealed by much shorter lived, ~7 Myr, bright $H_\alpha$ emitting, HII shells on the inner sides of the cocoons, ionized by the intense far ultraviolet emission of the Wolf Rayet stars before they blow off most of their mass, collapse and explode as SN Ic (van Dyk, et al., 1996; Anderson, et al., 2012).

The acceleration of cosmic ray protons in a superbubble was at last confirmed (Ackermann et al., 2011) by *Fermi* detection of the signature high energy gamma rays from the decay of pions produced by cosmic ray interactions in the cocoon around the largest nearby superbubble, the Cygnus Superbubble (Figure 3), even before such emission was seen from any of those lone core collapse supernova remnants in the interstellar medium interacting with close by clouds.

The Cygnus Superbubble, one of the brightest in the Galaxy, is fed by at least ten OB associations and smaller clusters, and it has been growing steadily for at least 20 Myr (Comeron et al., 2016, and the references therein). In Cygnus OB2, the youngest and by far the biggest of these, some 55 O stars of 15 $M_\odot$ or more have currently been identified (Wright et al., 2015), plus at least 88 early B0-B2 star supernova progenitors of 15 to 8 $M_\odot$. Most of the O stars in Cyg OB2 were produced between 1 and 7 Myr ago, possibly peaked around 4 to 5 Myr ago (Wright et al. 2015). Since the lifetimes of the most massive of these stars are only ~3 Myr (e.g. Schaller et al., 1992; Chieffi & Limongi 2013), a number of these stars have already exploded and others have lost significant mass in their winds, so their mass distribution has evolved considerably.

Nonetheless, studies show (Wright et al. 2015) that the present evolved population of OB2 is still consistent with expected evolution of an original Salpeter mass function with an integral index of -1.3. Therefore, considering an essentially complete sample of 36 O stars of 20 to 40 $M_\odot$ (Wright et al. 2015), all of whose lifetimes of >6 Myr exceed their age, we find that they would have made up 44% of all the original O stars, and thus we estimate that there were ~82 O stars in the initial population. Also based on an initial mass function ratio of the O stars (15-120 $M_\odot$) to early B stars (8-15 $M_\odot$) of 0.70, we expect that there were initially ~117 B0-B2 stars. Together they should produce a total of ~199 supernovae over their full explosive lifetime of ~32 Myr, occurring at an average rate of ~6 SN/ Myr in Cyg OB2

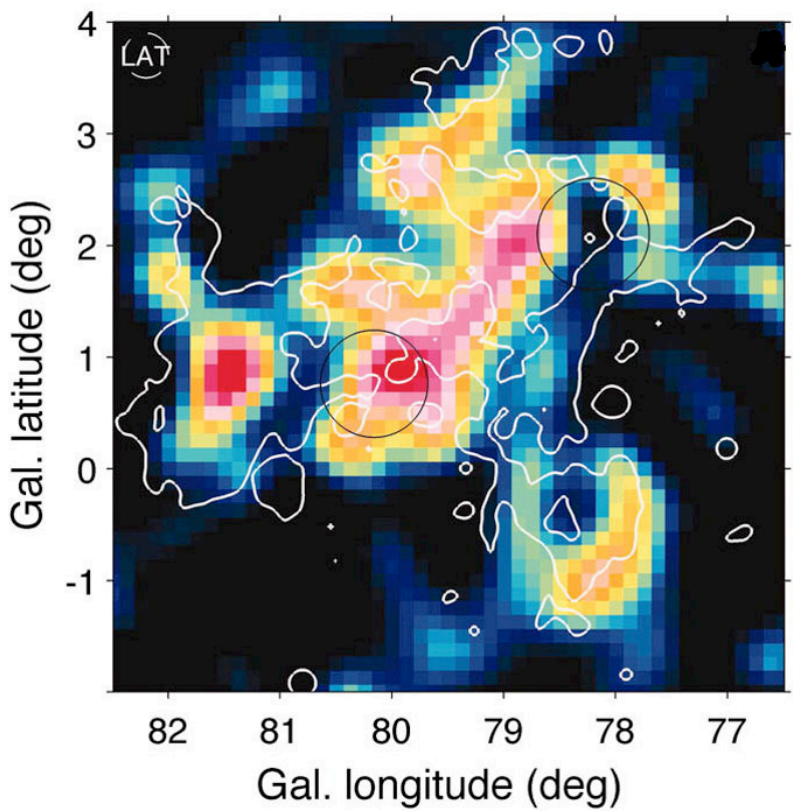

Figure 4. Illuminated by cosmic-ray produced, high-energy 0.1-100 GeV gamma rays, the Cygnus Superbubble is a giant, meandrous cosmic-ray filled cavity of ~$2\times10^5$ pc$^3$, with a diameter of ~35 pc reaching about ~100 pc, or ~4°, across, and a comparable distance along the line of sight that links OB associations in the nearby ~1.4 kpc Cygnus X region. The white contours outline the infrared emission from the surrounding cocoon (Ackermann et al., 2011).
.
The number of supernovae that have already occurred in Cyg OB2 can be estimated in two different ways. First, taking this average rate, which is only weakly dependent on age for this mass function, together with extended star formation in Cyg OB2 from 1 to 7 Myr ago with a possible peak 4-5 Myr ago, and the onset of supernova explosions in the most massive stars ~3 Myr later, we would expect that at least ~6 of the most massive O stars have already exploded as SN Ic within the last 1 to 2 Myr. Independently, from the 36 O stars observed between 20 and 40 $M_\odot$, the Salpeter mass function would imply that there were originally ~18 with masses >40

M$_\odot$, but only 10 are now observed (Wright et al. 2015). Since all but 2 of these 10 have masses <52 M$_\odot$, this leaves a conspicuous deficit of 8 stars in the mass range from 52 to 92 M$_\odot$, where *no* O stars are now found. There are, however, 3 former O stars now in their Wolf Rayet phase, which could account for part of the deficit, but even that still strongly suggests that at least 5 of the missing stars have already become supernovae. This is in good agreement with the other estimate of ~6 supernovae.

Although none of these supernovae would have left a detectable radio remnant in the low density superbubble, a few of those of ~50 to ~80 M$_\odot$ may have left neutron star remnants (Woosley & Weaver 2002). One $1.15 \times 10^5$ year old pulsar, thought to lie within Cyg OB2, has in fact been discovered (e.g. Wright et al. 2015).

In addition, of course, there have already been many supernova explosions of the O and B stars in nearly a dozen other older clusters within the Cygnus superbubble with ages of as much as ~20 Myr. A total of ~330 OB star (>3 M$_\odot$) candidates have been identified (Le Duigou & Knodlseder 2002) within the superbubble of which as many as ~1/4, or ~80, might be O and early B star supernova progenitors (>8 M$_\odot$), based on a Salpeter mass function, exploding at a rate of as much as ~3 SN/Myr over their lifetimes. Moreover, the oldest and most deeply investigated of these other clusters has only recently been recognized (Comeron et al., 2016), not from its O stars which have already exploded but from 7 B star, red supergiants in a cluster slightly offset, but overlapping with Cyg OB2. These stars, the most massive of the B stars at ~10 to 15 M$_\odot$, were born some ~15 to 20 Myr ago and they have become some of the brightest in the Galaxy in the infrared, as they enter their red supergiant phase before they finally explode as SN II. Their much older cluster, about 1/6 the size of Cyg OB2 (Comeron et al., 2016), has added ~1 SN/Myr to the Cygnus Superbubble for the last 17 Myr and will continue to do so for roughly another 15 Myr.

Thus, the total supernova rate in the Cygnus Superbubble is expected to be at least ~ 6 to 9 SN/Myr and over the typical superbubble lifetime of ~50 Myr this implies that it will produce about 500 core collapse supernovae, placing it indeed well above the superbubble average of ~300 supernovae. Their cosmic ray generation of about ~(2 to 3) $\times 10^{37}$ erg/s, assuming ~$10^{50}$ erg/SN, seems fairly well constrained. But the total energy of those cosmic rays currently residing in the bubble depends on their average residence time. Recent model calculations (Butt & Bykov 2008; Bykov 2014) of cosmic ray acceleration in superbubbles suggest that their lifetimes there are around 1 to 2 Myr. That would suggest that the Cygnus Superbubble currently contains somewhere between ~0.6 and 2 $\times 10^{51}$ ergs in freshly accelerated cosmic rays.

The total energy in all the cosmic rays presently filling the Cygnus Superbubble can also be determined observationally from the *Fermi* measurements (Ackermann et al., 2011) of the high energy gamma-ray luminosity of ~$10^{35}$ erg/s and the mass of the gas in which the cosmic rays interact. The *Fermi* group has estimated the mass of the HII gas in the cocoon shell from measurements of the column density along lines of sight through the bubble in excess of the surrounding background and foreground values measured outside. Depending on the assumed electron fraction, they found a cocoon HII mass of 0.32 to 1.6 $\times 10^5$ M$_\odot$. Calculating the superbubble cosmic ray energy interacting with such a range of HII mass that would be needed to produce the observed gamma ray luminosity, they found that all that was required was 1.3 to 6.5 $\times 10^{49}$ ergs, or only between ~10% and ~50% of that generated by a single core collapse supernova.

As we saw, this is only a fraction of the total cosmic ray energy being generated in the Cygnus superbubble, yet it is still quite compatible with what might be expected. For in such a turbulent region as that in the interface between the hot supernova-driven bubble and its swept up and compressed HII shell, we might expect that the cosmic rays diffuse into just the innermost edge of the shell and that those cosmic rays would amount to only a fraction of the collective output of many supernovae.

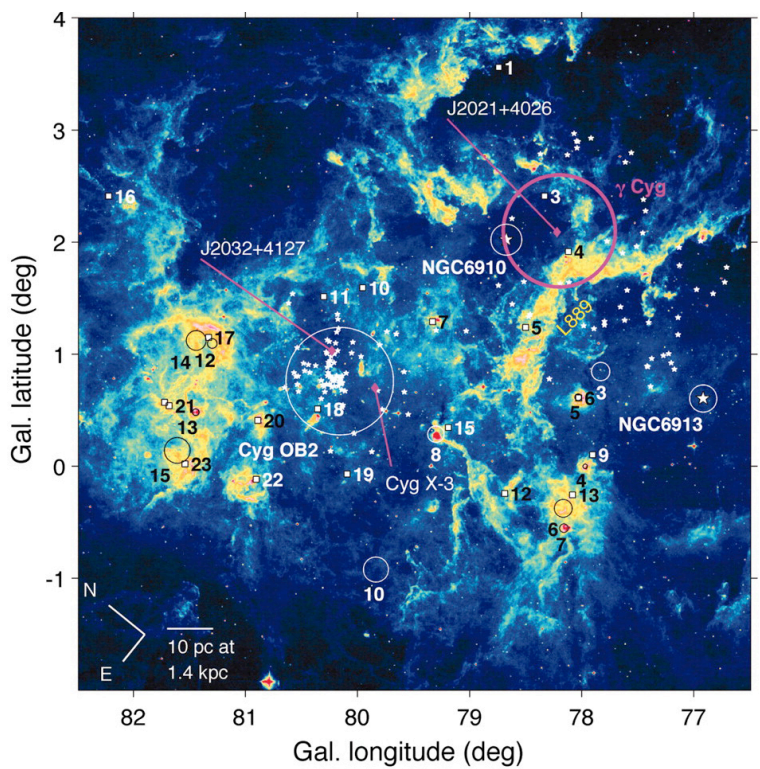

Figure 5. The Cygnus Superbubble lies in a tumultuous star forming region of Cygnus molecular cloud complex in which at least ten major OB associations have been identified, the youngest and brightest of which is Cyg OB2 in the heart of the bubble just left of center, surrounded by numerous smaller OB clusters (numbered circles) contributing to or engulfed by the superbubble. (Ackermann et al., 2011).

For, if as expected the combined $\sim 10^{51}$ ergs of cosmic rays from roughly $\sim 10$ supernovae from Cyg OB2 are trapped within the bubble for a couple million years in the very highly turbulent plasma generated by the supernovae, these cosmic rays could generate the observed gamma ray emission, diffusing into and interacting with a mass of only $1.6 \times 10^3$ $M_\odot$, or only a thin $\sim 1$ to 5% of the innermost edge of the surrounding cocoon mass, that could produce the high energy gamma ray luminosity measured by *Fermi*. Clearly detailed calculations of the superbubble-cosmic ray interactions with its surrounding cocoon are needed.

As has been frequently pointed out (e.g. Axford, 1981; Bykov & Fleishman, 1992: Bykov 2014), supernova shock acceleration is most efficient in the hot, low density plasma of the superbubbles, where the bulk of Galactic supernovae occur, since their shocks can propagate throughout the ejecta of the previous supernovae and winds without sweeping up enough surrounding gas to radiate away much of their energy, as many do in the warm, much denser phases of the interstellar medium. Those supernovae in superbubbles can also accelerate nuclei more effectively from the tenuous, supernova-enriched dust and gas with far lower competing energy losses than in the denser phases. In addition, over the ~50 Myr lifetimes of spatially and temporally clustered supernova generation in superbubbles their shocks can further accelerate lingering cosmic rays from earlier supernovae.

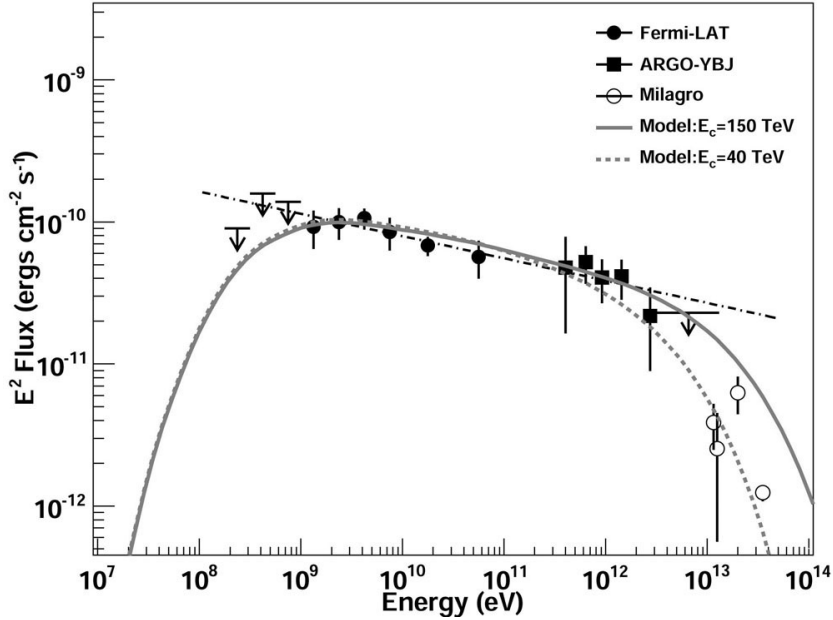

Figure 6. The spectrum of the ~100 MeV to ~30 TeV gamma rays produced by cosmic ray protons in the Cygnus Superbubble, measured by *Fermi* (Ackermann et al., 2011), *ARGO* (Abdo et al., 2009) and *Milagro* (Bartoli et al., 2014), showing the pion-decay signature below ~1 GeV, and the power-law acceleration index of -2.2 above extending all the way to ~3 TeV before breaking to the escape spectrum above, a factor of 100 to 1000 times higher in energy that in the single core collapse supernova remnants IC443 and W44 in very dense molecular gas and nearly a factor of ~10 higher than Tycho's remnant in the lower than average density gas (from Bartoli et al., 2014).

So the extensive superbubble acceleration processes also appear (Bykov & Toptygin, 2001; Ferrand & Marcowith, 2010) to be able to account for both the acceleration and the escape spectra of the locally measured cosmic rays with the simple -2.7 power-law energy spectra with constant nuclear abundance ratios up to the so called "knee" above $10^6$ GeV. There the proton spectra breaks (Horandel, 2013) at $E_p \sim 4 \times 10^6$ GeV, followed successively at higher energies by the less abundant helium and heavier nuclei breaking at $E_Z \sim E_p Z$, where Z is their nuclear charge, all the way up to U (Z ~92) at ~$4 \times 10^8$ GeV as the mean nuclear mass of the remaining cosmic rays increases. At the same time the further acceleration by the multiple supernova shocks help smooth the overall spectral index to about -3.1 above the knee on up to the so-called "ankle" at ~ $10^9$ GeV.

Moreover, stochastic acceleration in the turbulence generated by the massive star winds in the superbubbles may further push the cosmic rays to energies of at least $10^9$ GeV (Bykov & Toptygin, 2001), approaching the maximum measured cosmic ray energies of $10^{11}$ GeV.

## 4. Cosmic Ray Composition, Supernova Grain Injection and Superbubble Mixing

Astronomical observations have shown that superbubbles are where most of the Galactic cosmic rays are accelerated by supernova shocks. But direct measurements of the elemental and isotopic composition of our local cosmic rays as a function of energy have revealed much more on the processes within the bubbles. The local composition reveals the secondary nuclear spallation products produced during their propagation from their sources through the Galaxy to the solar vicinity and from which the original source composition can be determined.

That cosmic ray source composition together with calculations of the nucleosynthetic yields reveals the details of how newly synthesized nuclei in the massive star winds and supernova ejecta are mixed with the gas and dust in the bubble cores from which the cosmic rays are accelerated. These composition measurements also reveal the separation of the ejecta into high velocity dust grains and hot gas that further modify the mix by selectively injecting the nuclei into supernova shocks as suprathermal ions, biased by their atomic mass and their solid or gaseous state, as they are finally accelerated to cosmic ray energies in superbubbles. The abundances of radioactive $^{59}$Ni and $^{60}$Fe in the superbubble source mix and the local cosmic rays, set time constraints between the synthesis, acceleration and propagation.

First, the abundances of secondary nuclei in local cosmic rays show the effects of spallation interactions during their propagation through the Galaxy, and reveal the energy-dependent path length in g/cm$^2$ of interstellar gas through which they have passed. That path-length distribution is derived (e.g. Engelmann et al., 1990; Obermeier et al., 2012) from the energy dependence of secondary boron to primary carbon abundance ratio, B/C, which is ~ 0.35 at 1 GeV/nucleon, and corresponds to a path length of ~11 g/cm$^2$ at that energy. Correcting the locally meaasured cosmic ray abundances for the increase of all the secondary nuclei and reduction of the primaries, expected from this path-length dependence has been used to determine their original source abundances.

The energy dependence of this path length which has a power law index of around -0.5 (Obermeier et al., 2012) further implies that the locally measured spectrum with a power law index of -2.7 has evolved from an original source spectrum with an index of -2.2, which is quite consistent with that expected for supernova shock acceleration in superbubbles, and determined from the Cygnus Superbubble, as discussed above.

Studies of the origin of Galactic B and its light neighbor Be, also showed (Reeves, Fowler & Hoyle 1970) that they are unique among all of the elements in the Galaxy, in that they are produced not by nucleosynthesis like the rest, but predominantly by spallation reactions of cosmic rays with the interstellar gas. There are two complimentary but distinct processes of BeB production. The most direct is spallation of energetic cosmic ray CNO and other nuclei by ambient interstellar H and He, whose BeB production rate is directly proportional to the supernova production rate of CNO. The complimentary process is the spallation of the slowly accumulating ambient CNO and heavier nuclei in the interstellar gas and dust by energetic cosmic ray protons and alpha-particles, whose BeB production rate is proportional to the accumulation rate of CNO in interstellar medium, which is in turn proportional to the integral of the CNO production.

Estimates of the present BeB production rate in our Galaxy (Reeves, Fowler & Hoyle 1970; Ramaty et al., 2000) suggest that the two processes are roughly equal in the current, well-evolved interstellar medium. But extensive measurements tracing the evolution of the BeB/H abundance ratios in the atmospheres of stars in our Galaxy from the very oldest born roughly ~10 Gyr ago, when the interstellar medium metallicity $Z/Z_\odot$ was $< 10^{-3}$ of the solar value to the present value of just 1.3 times solar, show (Duncan, Lambert & Lemke, 1992; Alibes, Labay & Canal, 2002) that over nearly the full range of stellar metallicities up to ~0.3 of solar the BeB production is directly proportional to the supernova production of CNO.

This requires that a constant fraction of the supernova ejecta CNO be accelerated to cosmic ray energies. The simplest acceleration process would be for the ejecta of a supernova to be accelerated by its own shocks (Lingenfelter, Ramaty & Kozlovsky 1998). But this is effectively ruled out by measurements (Wiedenbeck et al., 1999) of the cosmic ray $^{59}$Ni and $^{59}$Co, a radioactive parent-daughter pair that decays with a mean-life of $1.1 \times 10^5$ yr only by bound electron capture and cannot decay once the $^{59}$Ni is accelerated and its electrons are stripped off. Models of SNII, which account for 67% of all core collapse supernovae (Li et al., 2011), predict (Woosley & Weaver, 1995) that the Salpeter initial-mass-function weighted radioactive Ni makes up 60±27% of the freshly produced mass 59 nuclei. On the other hand, the measured cosmic ray value, corrected (Wiedenbeck et al., 1999) for spallation during propagation, is 18.2±2.3%. This would require the expected primary SNII fraction not to have been accelerated $<2 \times 10^5$ yrs after its production, which is clearly inconsistent with acceleration by the supernova's own shocks.

Most recently, a primary $^{59}$Ni fraction just below the local cosmic ray limit has been calculated with a selected rotational model for the more problematical SNIb/c core collapse supernovae of more massive, >25 $M_\odot$ stars (Neronov & Meynet 2016), assuming that they occur even in the most massive stars of 120 $M_\odot$, and are not effectively quenched at > 40 $M_\odot$ by black hole formation. There efficacy of such production is also clouded by further uncertainty in the overall $^{59}$Ni fraction because of the currently unknown contribution of the thermonuclear SNIa.

Such a minimum delay time of $>2 \times 10^5$ yrs from nucleosynthesis to acceleration, implied by the apparent lack of primary $^{59}$Ni in the local cosmic rays (Wiedenbeck et al., 1999), however, is actually quite consistent with expectations for a superbubble origin (Higdon, Lingenfelter & Ramaty 1998; Binns et al., 2008), where the ejecta of OB clustered supernovae mix with gas in the bubble and are accelerated by the collective shocks of the supernovae for many Myr. For a typical number of ~100 or less supernovae over the ~35 Myr active life of a single OB association discussed above, the mean delay time from nucleosynthesis in one supernova to the accelerating shock generation of the next supernova is $\sim 3.5 \times 10^5$ yr, while for an average of ~300 supernovae in the larger, merged superbubble around several OB associations with a combined life of ~50 Myr the mean interval is $\sim 1.7 \times 10^5$ yr. Despite the expected statistical scatter in individual delay times, these means are greater than the implied mean time for solely the SNII $^{59}$Ni, not counting its reduction by mixing of that eject with interstellar gas and the contributions of SNIa and Ib/c. Nonetheless, even the lack of a minimum time constraint would not argue against cosmic ray acceleration in superbubbles.

However, it is the primary cosmic ray source composition determined from the local cosmic rays that provides the greatest insight into the crucial processes of its injection and acceleration. This composition (Figure 7), as determined by Engelmann et al. (1990), shows that the cosmic ray source abundances are highly enriched in most heavy nuclei, i.e. atomic mass A > 4, compared to current solar and local Galactic abundances (Lodders, 2003). Nearly all of the heavy elements in the Galaxy have been produced by stellar nucleosynthesis (e.g. Timmes et al., 1995), and ejected in massive star winds and supernova explosions averaging ~18 $M_\odot$/SN with a metallicity ~10 times that of solar, which is then mixed into the interstellar medium primarily in the cores of superbubbles. Detailed analyses of the cosmic ray composition have provided further evidence of their superbubble origin. They have also revealed how the cosmic ray heavy nuclei are preferentially injected into the accelerating supernova shocks as suprathermal ions sputtered and scattered from high velocity dust grains of refractory elements in the supernova ejecta.

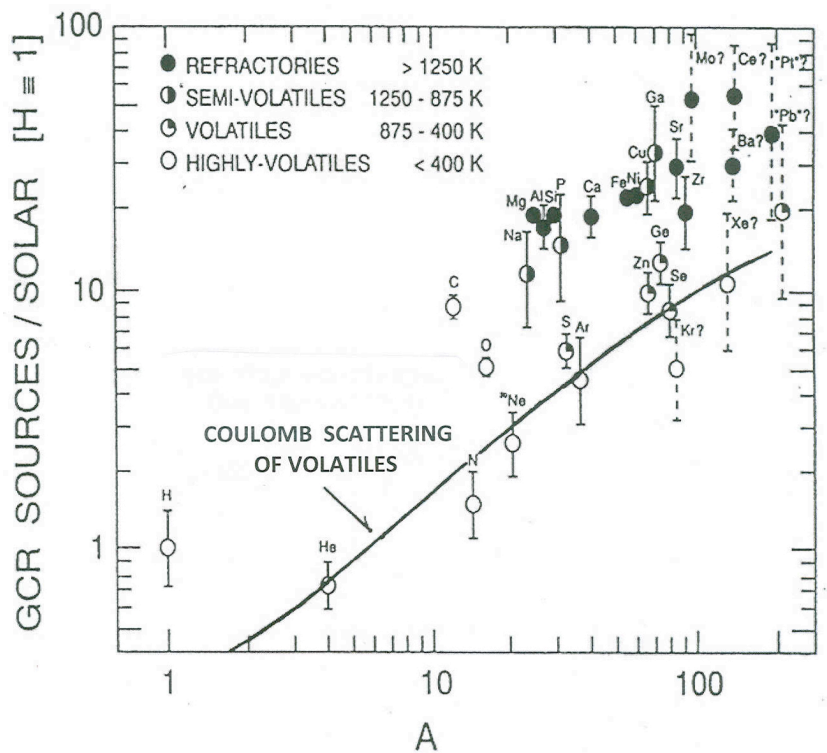

Figure 7. Galactic cosmic ray source versus solar composition shows the roughly 20X or more enrichment of the "refractory" elements in the cosmic rays, sputtered from fast, supernova-enriched dust grains, and the clear atomic mass, A, dependence of the "volatile" elements, expected from their Coulomb scattering off the fast grains. (Adapted from Meyer et al., 1997; Lingenfelter & Higdon, 2007)

Basically the cosmic ray source abundances (Figure 7) are enriched by a factor of 20 to 30 compared to the solar abundances in the so-called "refractory" elements. These refractory oxides (e.g. $MgO$, $Al_2O_3$, $SiO_2$, $CaO$, $Fe_3O_4$) appear (Matsuura et al., 2011) to rapidly, < 30 yr, condense in the adiabatically cooling ejecta of the core collapse SN 1987A to form high velocity, ~3,000 km/s, dust grains, once the co-moving plasma temperature drops below about 1250 K. This strong enrichment of the refractories clearly indicates that these nuclei were preferentially accelerated, because they were injected into accelerating shocks as suprathermal ions, produced in the breakup of fast refractory dust grains by sputtering in the ambient highly ionized gas in the shocks (Cesarsky & Bibring, 1981; Meyer et al., 1997; Lingenfelter et al., 1998), rather than being selected by their lowest ionization energy in the warm phases of the interstellar medium as was previously suggested.

Moreover, an atomic mass, A, dependence of the less enriched heavy volatile elements is clearly shown over their much wider range of measured masses. These elements, which remain in the gas at even much lower temperatures, have also been shown (Lingenfelter & Higdon, 2007) to be quite consistent with their being scattered out of the gas and selectively injected as suprathermal ions at the grain velocity by the strongly charge, or ~A/2, dependent Coulomb scattering (Sigmund, 1981) in those same interactions of the fast dust and gas. This sputtering cross section between two nuclei of masses a and A is proportional to $aA / (a^{2/3} + A^{2/3})^{1/2}$, which for the heavy elements, A > 4, interacting with protons, a = 1, reduces to roughly $\sim A^{2/3}$.

The intermediate enrichments of O and C have also been shown (Lingenfelter & Higdon, 2007) to be consistent with the combined supernova and wind yields in both the refractory and volatile phases, with suprathermal injection by sputtering of both the O in the refractory oxides and C in graphite in the fast grains by ambient hydrogen, together with grain scattering of the volatile O and C remainders in the ambient gas.

The mixing of freshly synthesized material with the surrounding interstellar, $Z_{ism} \sim 1.3\ Z_\odot$, gas and dust has been quantified by extensive measurements of local cosmic ray elemental and isotopic ratios, since they are accelerated out of that mix with the same atomic mass dependent Coulomb scattering dependence and injection. These studies (Higdon & Lingenfelter, 2003) began with the measured cosmic ray $^{22}Ne/^{20}Ne$, which is fully 5.3 times that of solar system ratio (Binns et al., 2005). Allowing for large uncertainties in the combined yields of these isotopes in Wolf Rayet winds and core collapse SNIb/c and SNII, This isotope ratio implies a robust mixing ratio of 18±5% massive star ejecta with 85±5% interstellar gas. Similar massive star mixing ratios of about 20% have since been found (Binns et al., 2005, 2006, 2008), using more recent Wolf Rayet models, for the $^{22}Ne/^{20}Ne$ and the other two largest cosmic ray isotopic deviations from solar system values, $^{12}C/^{16}O$ and $^{58}Fe/^{56}Fe$.

Other studies (Lingenfelter et al., 2003; Lingenfelter & Higdon, 2007) of significant cosmic ray elemental enrichment ratios, C/Fe, O/Fe, Si/Fe, and ThU/Pt group from supernovae also all suggest ejecta mixing ratios close to 20%. The latter calculations (Lingenfelter et al., 2003) of the radioactive actinide/Pt group abundance ratio not only predicted a cosmic ray value of 0.027±0.005, which was quite consistent with the latest balloon measurements (Combet et al., 2005; Donnelly et al. 2012) of ~0.032. But the predicted mean ratio in the dust and gas of the superbubble core is 0.029±0.005, or about twice the present interstellar value, is in fact consistent with the protosolar ratio of ~0.23 from which the sun was formed 4.5 Gyr ago (e.g. Fowler 1972).. This clearly suggests that the Sun too was formed in a superbubble like most other stars, and not in some chance close encounter with a random supernova, as had been previously suggested.

High energy measurements (Ahn et al., 2010) of the abundance ratios of cosmic ray He, C, N, Mg, Si and Fe relative to O also show that the same mixing ratio holds constant over at least 4 decades of energy from ~100 MeV/nucleon up to ~4 TeV/nucleon. Thus these local cosmic ray measurements all suggest an essentially constant cosmic ray source component of ~20% high metallicity ejecta of freshly synthesized nuclei over a wide energy range for the nominal ~ 20 Myr lifetime of the current local < 1 kpc cosmic rays. And, as we saw, analyses (Ramaty et al., 1997; Alibes et al., 2002) of old halo star abundance ratios of Be, produced primarily by the spallation of cosmic ray CNO, give a similar constant ratio of ~25% in the far older, more distant cosmic ray composition extending out to ~ 10 kpc and back ~10 Gyr, when the average metallicity of the interstellar medium was as low as, $Z_{ism} \sim 10^{-3}\ Z_\odot$.

Recently, cosmic ray abundances measurements on the *ACE* spacecraft (Ogliore et al., 2009) and the Antarctic long-duration balloon flights *TIGER* and *Super-TIGER* (Rauch et al., 2009; Murphy et al., 2016) not only further confirmed the cosmic ray source mixing ratio with a best-fit fraction of 19% freshly synthesized massive star ejecta and 81% old interstellar gas and dust, but they also quantified the subsequent elemental injection biases that further modified the cosmic ray composition. In a self-consistent analysis they determined the best-fit values of both the mass-dependent injection of the source mix into the supernova shocks as suprathermal ions sputtered off fast refractory grains by ambient H and He, and the resulting, strong overall preferential injection of refractory elements over and volatiles (Figure 8).

The cosmic ray atomic mass dependence of the refractory and volatile element abundances relative to that of the source mix are modeled by power laws with best-fit exponents of 0.58±0.07 and 0.63±0.12, respectively. These values are inconsistent at 2-σ or more with an $(A/Q)^{0.40}$, mass-over-charge, dependence expected (Meyer et al., 1997) from a rigidity dependent acceleration efficiency. As we saw above, however, this abundance dependence is quite consistent with an ~$A^{2/3}$ power law expected (Sigmund 1981; Lingenfelter and Higdon 2007) from the Coulomb scattering of volatile elements in the surrounding gases by high velocity refractory elements in the grains in the supernova ejecta. Similarly the refractory power-law dependence is also consistent at just over 1-σ with the same cross section dependence for the sputtering of refractory elements from the high velocity grains interacting by the ambient gas through which they pass.

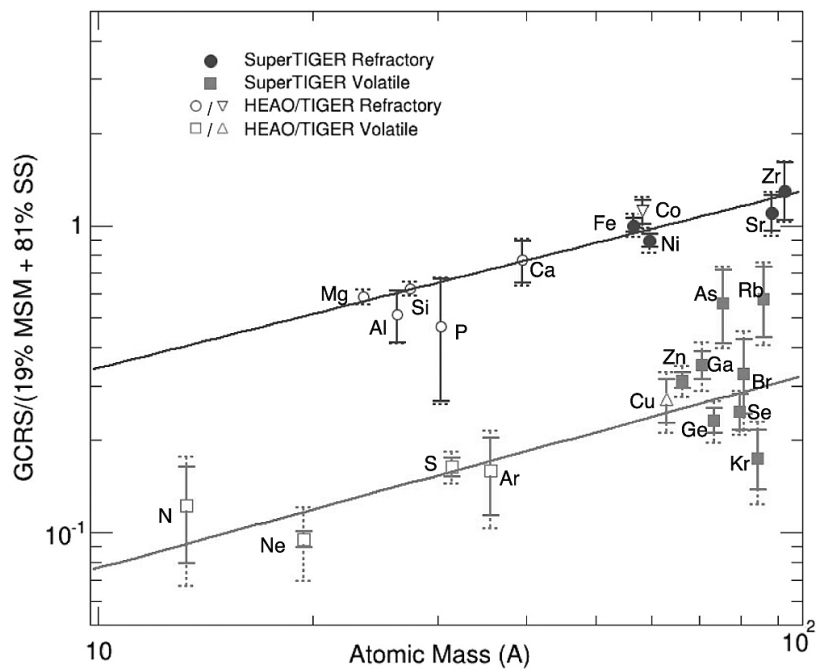

Figure 8. The atomic mass dependence of the relative cosmic ray source refractory and volatile element abundances with best-fit power-law exponents of 0.58±0.07 and 0.63±0.12, respectively, together with the measured 4:1 injection bias for refractory versus volatile elements, all compared here to that for the best-fit source mixing ratio of 19% massive star ejecta and 81% solar system abundances rather than unenriched solar abundance shown in Figure 7. (from Murphy et al. 2016, Fig. 9).

The sputtering and scattering of refractory and volatile elements to supernova ejecta grain velocities provides the suprathermal ions selectively injected into the accelerating shocks. This further enriches the heavier ions in the cosmic rays relative to the cosmic ray source mix. Moreover, as these measurements clearly show the refractories are enriched by a factor of 4 compared to the volatile elements of similar atomic mass, because the grains alone carry the energy that produces the suprathermal ions needed for effective injection, while the volatile elements remaining in the gas lose their energy much faster through plasma interactions and thermalization.

Perhaps most important, the sputtering of a refractory atom from a grain only requires breaking a relatively low energy chemical bond to unleash a suprathermal ion without any significant extra acceleration, while the scattering of volatile atoms in the gas up to suprathermal energies requires a relatively much more energetic and less likely interaction to accelerate volatile ions suprathermally. Therefore the injection of volatile ions accelerated out of the ambient gas should be significantly less likely than the injection of refractory ions simply stripped off high velocity grains, as is clearly seen. This striking difference between the refractory and volatile cosmic ray abundances clearly demonstrates the critical importance of fast grain sputtering in the injection of suprathermal ions into shock acceleration.

Extensive measurements by *CREAM* (Ahn et al., 2010) and *AMS-02* (Aguilar et al., 2015, 2016), of the cosmic ray He/p and C/p flux ratios from 10 GeV to a few TeV also reveal a rigidity dependent hardening of both the He and C spectra with respect to H with a power law index of +0.08±0.02. This divergence, not expected from diffusive shock acceleration, has been attributed to a wide range of mechanisms, e.g., spallation during propagation (Blasi & Amato, 2012), acceleration out of stellar winds (Biermann et al., 2010), and out of reverse shocks in SN Ia versus SNII (Ptuskin et al., 2013), all in individual supernova remnants, as well as in a supernova ejecta enrichment gradient in superbubbles (Ohira et al., 2016). But, aside from the fact that single supernovae in the interstellar medium make up only ~15% of Galactic supernovae, which would require a huge effect to affect the whole cosmic ray spectrum, all of the processes proposed for them are also arguably (Ohira et al., 2016) inconsistent with other measurements, e.g. B/C, or rely on unknown mechanisms, e.g. the suppression of either H or He acceleration by the reverse shocks in the winds and the forward shocks in the interstellar medium.

Superbubbles, on the other hand, which with greater acceleration efficiency account for over 75% of the cosmic ray spectrum (Higdon & Lingenfelter, 2005), are naturally expected to have a higher ejecta enrichment of the bubble gas due to the concentration of the supernovae at the core of the OB association where the strong initial shocks accelerate cosmic ray to their highest energies with the hardest spectra, while the subsequent expanding and weakening of the shocks generate ever softening spectra with decreasing ejecta enrichment down to lower energies, producing the observed divergence between the ejecta dominated heavy nuclei, such as He and C, and the interstellar-medium-dominated H (Ohira et al., 2016). These measurements can also now help quantify the timescale of this initial ejecta mixing in the bubble gas, and hopefully shed new light on the mean mixing ratio.

All of these cosmic ray abundance ratios are quite consistent with their being primarily accelerated out of a common characteristic mixed medium in the cores of superbubbles consisting of roughly 20% by mass of freshly synthesized material in the ejecta and winds of massive stars mixed with ~80% older ambient interstellar medium. It should be emphasized, that the relative proportions of all the massive star wind and supernova heavy nuclei with A > 4 compared to their counterparts in the older interstellar medium is much larger than these bulk weights, which also include all of the H and He, would seem to imply. For the metallicities of these components are very different with that of the massive-star heavy nuclei having a mass fraction of ~0.2, being ~ 10 times that fraction of 0.02 in the nominally solar interstellar medium over the past 5 Gyr. Therefore, with this common superbubble mixing ratio of ~20% to ~80% (Lingenfelter & Higdon, 2007) in this superbubble mix the mass fraction of the fresh metal is 0.04 compared to 0.016 in old metals. Thus, there are actually 2.5 times as many the freshly synthesized heavy nuclei as there are older heavy nuclei, and these *freshly synthesized nuclei make up fully ~70% of all the heavy nuclei accelerated to cosmic rays in the bubbles*. Since this superbubble metallicity is ~2.4 times that in the current interstellar medium, the superbubble contribution to the heavier, A > 4, cosmic rays is even bigger, > 85%, than their > 70% contribution to the cosmic ray protons and He.

The cosmic ray composition, in fact, provides by far the most reliable way of measuring the supernova metal enrichment in the superbubbles surrounding star formation regions. For even though the X-ray spectra of the hot, > $10^6$ K, plasmas in superbubbles have been measured, they cannot detect most of this enrichment, because the major metals are mostly refractory and are in the grains instead of the plasma.

All of these analyses (Lingenfelter & Higdon, 2007) have also included the contributions to the cosmic ray metals by the thermonuclear SNIa, which account for the other 19% of Galactic supernovae (Li et al., 2011) and occur predominantly in the general interstellar medium. Their ejecta composition (Nomoto et al., 1984) is radically different from that of the core collapse supernovae, most notably with Fe making up ~3/4 of the total ejecta mass since their core Fe does not collapse into compact neutron stars, and for the same reason there is virtually no contribution to heavy nuclei from r-process nucleosyntheses. But, since we have no measurements indicating the injection biases or acceleration efficiency of their ejecta, we have simply added the unbiased composition of their ejecta to the cosmic ray mix.

Most recently, a promising new $^{60}$Fe/$^{56}$Fe nucleosynthesis-clock has been measured (Binns et al., 2016) that sets an upper limit on the time between nucleosynthesis, acceleration and propagation, which compliments the $^{59}$Ni lower limit. Radioactive $^{60}$Fe is predominantly synthesized in core-collapse supernovae and it decays by $\beta^-$ decay with a mean life of 3.78 Myr. Mixed with interstellar gas and dust in superbubbles, the $^{60}$Fe can be accelerated to cosmic ray energies. That produced within an estimated ~600 pc of the sun can reach the solar vicinity before they decay, and they have now been detected (Binns et al., 2016) at $^{60}$Fe/$^{56}$Fe of (4.4±1.7) x$10^{-5}$. This at last should allow us to study local cosmic ray propagation from all the nearby OB associations and their superbubbles (de Zeeuw et al., 1999), not only identifying the major sources, but placing significant new constraints on their diffusion mechanism.

## 5. Summary and Implications

We have seen that extensive astronomical observations of star forming regions and detailed measurements of the very unusual cosmic ray composition have provided compelling evidence that over 75% of the Galactic cosmic rays from about 0.1 GeV up to the "ankle" at around $10^9$ GeV are accelerated by supernova shocks out of highly ejecta-enriched dust and gas by spatially and temporally clustered bursts of core collapse supernovae, SNII and SNIb/c, in superbubbles formed around massive OB associations, while most of the remainder of the cosmic rays are accelerated by the shocks of single thermonuclear supernovae, SNIa, scattered randomly throughout the interstellar medium.

In particular, the recent high energy gamma ray observations of both individual supernova remnants and those forming superbubbles have power-law spectra with exponents of ~ -2, consistent with diffusive shock acceleration, as do the cosmic ray source spectra implied by direct measurements of the local cosmic rays from both protons and heavier nuclei corrected for propagation. Analyses of recent compositional measurements of the local cosmic rays, also show that their heavy A > 4 nuclei, which are over 20 times more abundant compared to protons than the interstellar medium, come from a mix of ~20% recent supernova ejecta with a high metal fraction of ~0.2, and ~80% older, ~0.02 ambient gas in superbubbles. Thus supernova ejecta makes up the major fraction, >70%, of the cosmic ray metals accelerated in superbubbles.

In addition, these analyses reveal how these dominant metals, which have mostly condensed in the adiabatically cooling ejecta as high-velocity refractory grains, are sputtered off by collisions with the shocked gas and injected as suprathermal ions into the cosmic-ray accelerating supernova shocks. This further favors the fast grain nuclei, compared to the refractory nuclei in the ambient interstellar grains and all of the volatiles in both the ejecta and the surrounding gas, enhancing the fast refractory grain nuclei by a factor of ~4 over the volatiles, as directly measured in the local cosmic rays. Since the sputtering cross section is roughly proportional to $\sim A^{2/3}$, this also greatly enriches the cosmic rays in heavier metals. All together these processes account for the enormous enrichment of these heavier nuclei in the accelerated cosmic rays.

Moreover, the cosmic-ray produced pion decay emission from the core collapse supernovae has now been clearly seen in the *Fermi* gamma-ray images not only of the nearby giant cosmic-ray-filled Cygnus Superbubble with an estimated ~200 OB star supernova progenitors, but also in some 30 to possibly 44 more distant individual supernova remnants in the interstellar medium, which are interacting with dense gas from molecular clouds and have been assumed to be from core collapse supernovae because of this association. Indeed, these latter are most likely some of the superbubble fringe from OB clusters too small to form a superbubble, but still close enough to interact with dense gas in their parent molecular clouds.

The SN Ia thermonuclear supernovae which are expected to produce the bulk of the single supernova remnants in the interstellar medium, have so far produced only about half a dozen cosmic ray produced pion decay gamma ray emitting remnants in interstellar gas of only average density, but all from those in the limited population of young remnants mostly <1 kyr old.

Finally, the source of the very highest energy cosmic rays above the ankle at $\sim 10^9$.GeV is still unknown (e.g. Olinto, 2013).

These new observations and analyses provide a more comprehensive framework of cosmic ray acceleration environments in which we can now undertake more detailed models of both supernova shock and turbulent stochastic acceleration and propagation in the unusual environment of superbubble cores. We can also now explore more realistic 3-D models of cosmic ray propagation from the more detailed temporal and spatial distribution of cosmic ray sources in the giant Cygnus Superbubble and the other smaller known ones nearer by to better assess their individual contributions to the local cosmic rays that we directly measure.

Acknowledgements

I am particularly indebted to two anonymous reviewers for very thoughtful comments and clarifications that have greatly improved this paper.